# Mismatch study of C-ADS main linac*

MENG Cai[1)](孟才), TANG Jing-Yu(唐靖宇), PEI Shi-Lun(裴士伦), Yan Fang(闫芳)

Laboratory of Particle Acceleration Physics & Technology,
Institute of High Energy Physics, Chinese Academy of Sciences, Beijing 100049, China

**Abstract** The ADS accelerator in China is a CW (Continuous-Wave) proton linac with 1.5 GeV in beam energy, 10 mA in beam current, and 15 MW in beam power. To meet the extremely low beam loss rate requirement and high reliability, it is very important to study the beam halo caused by beam mismatch, which is one major source of beam loss. To avoid the envelope instability, the phase advances per period are all smaller than 90 degree in the main linac design. In this paper, the results of the emittance growth and the envelope oscillations caused by mismatch in the main linac section are presented. To meet the emittance growth requirement, the transverse and longitudinal mismatch factors should be smaller than 0.4 and 0.3, respectively.
**Key words** Mismatch, C-ADS accelerator, emittance, beam loss
**PACS** 29.27.Bd, 29.27.Fh

## 1 Introduction

The ADS accelerator in China (C-ADS) is demanded to deliver a CW proton beam with 1.5 GeV in energy and 10 mA in current [1]. It is composed by two major accelerating parts: the injector and the main linac. The main linac is designed to boost the beam energy from 10 MeV up to 1.5 GeV with four accelerating sections, which the lattice structures of are shown in Fig. 1. The solenoid focusing is applied in the two spoke cavity accelerating sections and the triplet focusing is used in the two elliptical cavity accelerating sections.

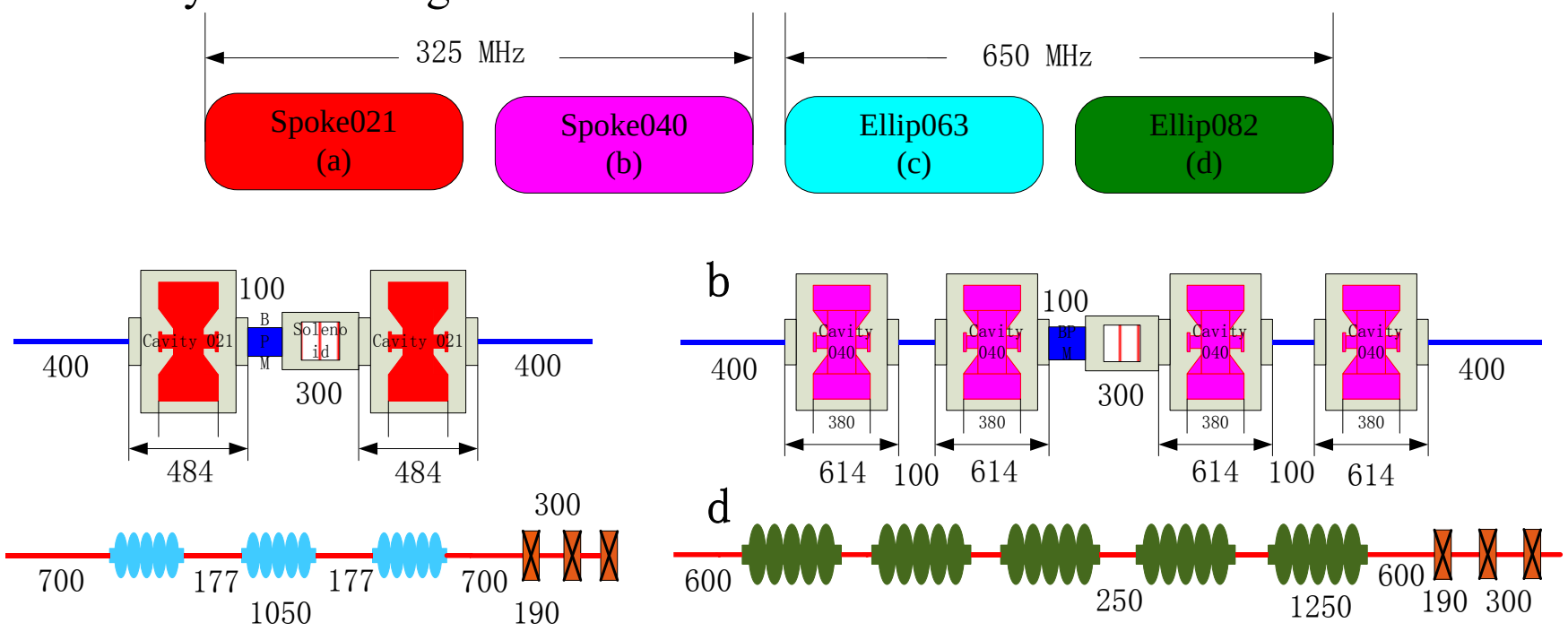

Fig. 1 Schematic view of the lattice structures for the main linac sections.

Beam loss rate of 1 W/m is widely used in the high-power proton accelerator, which is mainly determined by the hands-on maintenance requirement. To meet the extremely high reliability and availability for C-ADS accelerator, it is very important and imperative to study beam loss mechanism. Beam halo caused by mismatch is one major source of beam loss, where the halo particle can be lost on the walls of the beamline structures. The impact of misalignment errors and field errors is also very important and has been studied [2].

The emittance growth due to the mismatch should be carefully considered in the ADS accelerator where the beam loss is concerned. With the presence of nonlinear components, the filamentation effect will lead to a real emittance dilution, and the betatron modulation has the similar impact. More than a filamentation, a mismatched beam can be unstable if the channel working point is not properly set. Some mismatched modes can exhibit an instability when the phase advance without space charge per focusing period is greater than 90° and the tune depression is low [3]. In this paper, the simulation results of the ADS main linac with beam mismatch are presented.

## 2 MISMATCH

If the ellipse in phase space of injected beam is not matched to the downstream focusing system, there will be additional oscillations of the rms beam envelopes, which produce a larger sized beam at some locations and a smaller sized beam at other locations. In this scenario, it is convenient to define a mismatch factor, which is a measure of the increase in the maximum beam size resulted from a mismatch. Suppose that the matched beam phase space ellipse is defined by

$$\gamma_m x^2 + 2\alpha_m xx' + \beta_m x'^2 = \varepsilon$$

and a mismatched beam phase space ellipse with same area is defined by

* Supported by the China ADS Project (XDA03020000) and National Natural Science Foundation of China (11235012)
1) E-mail: mengc@ihep.ac.cn

$$\gamma x^2 + 2\alpha x x' + \beta x'^2 = \varepsilon$$

Then the mismatch factor can be expressed as

$$M = \sqrt{\frac{\Delta + \sqrt{\Delta^2 - 4}}{2}} - 1 \tag{1}$$

where $\Delta = \beta_m \gamma + \gamma_m \beta - 2\alpha_m \alpha$ .

The rms emittance growth due to beam mismatch whatever the beam distribution can be given by [4]

$$\xi = \frac{\varepsilon_{rms,mismatched}}{\varepsilon_{rms,matched}} = \frac{1}{2}\left( \frac{\beta_m}{\beta} + \frac{\beta}{\beta_m} + \frac{\beta_m}{\beta}\left( \alpha - \frac{\beta}{\beta_m}\alpha_m \right)^2 \right) \tag{2}$$

and the total emittance growth is given by

$$\eta = \xi + \sqrt{\xi^2 - 1} = (1+M)^2 \tag{3}$$

In addition, filamentation of the particle distribution in phase space can cause emittance growth. Progress has been made in understanding halo production due to parametric resonances between single particle and the oscillating mismatched beam core. The mismatch of DC beams is described by two well-known eigenmodes [3]. For bunched beams, there are 3 eigenmodes [5], a pure transverse quadrupole mode

$$\sigma_{env,Q} = 2\sigma_t \tag{4}$$

and a high mode and a low mode which couple the transverse and longitudinal directions

$$\sigma_{env,H}^2 = A + B,\ \sigma_{env,L}^2 = A - B \tag{5}$$

with

$$A = \sigma_{t0}^2 + \sigma_t^2 + \sigma_{l0}^2/2 + 3\sigma_l^2/2$$

$$B = \sqrt{\left(\sigma_{t0}^2 + \sigma_t^2 - \sigma_{l0}^2/2 - 3\sigma_l^2/2\right)^2 + 2\left(\sigma_{t0}^2 - \sigma_t^2\right)\left(\sigma_{l0}^2 - \sigma_l^2\right)},$$

where $\sigma_t$, $\sigma_{t0}$, $\sigma_l$ and $\sigma_{l0}$ are full and zero current transverse and longitudinal phase advance. With smooth approximation, one can get the corresponding Eigen-solutions for quadrupole mode

$$\Delta a_x/a = -\Delta a_y/a = A_m \cos\left(\sigma_{env,Q}\, s/L + \phi\right),\ \ \Delta b/b = 0 \tag{6}$$

and for high and low mode

$$\Delta a_x/a = \Delta a_y/a = g_{H/L}\Delta b/b = A_m \cos\left(\sigma_{env,H/L}\, s/L + \phi\right) \tag{7}$$

where

$$g_{H/L} = \frac{\sigma_{t0}^2 - \sigma_t^2}{\sigma_{env,H/L}^2 - 2\left(\sigma_{t0}^2 + \sigma_t^2\right)} \tag{8}$$

$g_H$ is always positive and $g_L$ is always negative. This approximation is not valid for extensively elongated bunches.

According to above analysis, we use beam size differences to characterize mismatch for convenience, so the twiss parameters can be represented by following expression for one mismatch factor $M$, which is signed "+".

$$\beta = (1+M)^2 \beta_m,\ \alpha = (1+M)^2 \alpha_m,\ \Delta a/a = M \tag{9}$$

In addition, there is another case shown by following expression, which is signed "-".

$$\beta = \beta_m/(1+M)^2,\ \alpha = \alpha_m/(1+M)^2,\ \Delta a/a = -M/(1+M) \tag{10}$$

For convenience, we define quadrupole mode mismatch factor as $M_Q = \Delta a/a$ and high and low mode mismatch factor as $M_{H/L} = \Delta b/b$ in this paper.

## 3 Multiparticle simulation results

In the multi-particle simulation, one track $5\times10^6$ particles with 5σ&6σ Gaussian distribution. In the mismatch study, the transverse and longitudinal rms emittance growths are controlled to be smaller than 25% and 20% respectively, which is the same emittance growth level caused by errors.

### 3.1 Filamentation effect

Firstly, we study the filamentation effect with space charge lead to emittance growth. For the longitudinal mismatch, the rms emittance growths with different mismatch factor are shown in Fig 2. The simulation results agree well with the theoretical ones given by Eq. 2. Transverse emittance growth is caused by space charge effect. Considering the symmetry in transverse of solenoid focusing structure, we only simulated the dynamic results with mismatch in x plane. Due to the x/y coupling caused by solenoid in transverse direction and space charge effect, the emittance growth in three planes shown in Fig 3. The difference of emittance growth in two transverse planes is because of the difference of matched beam size at Spoke040 section, which is caused by the field asymmetry of Spoke040. From the simulation results, there are beam losses when transverse mismatch factor is larger than 0.8. For these cases, the transverse mismatch factor should be smaller than 0.5 to meet the designed requirement and the longitudinal mismatch factor should be smaller than 0.3, which are shown in Table 1.

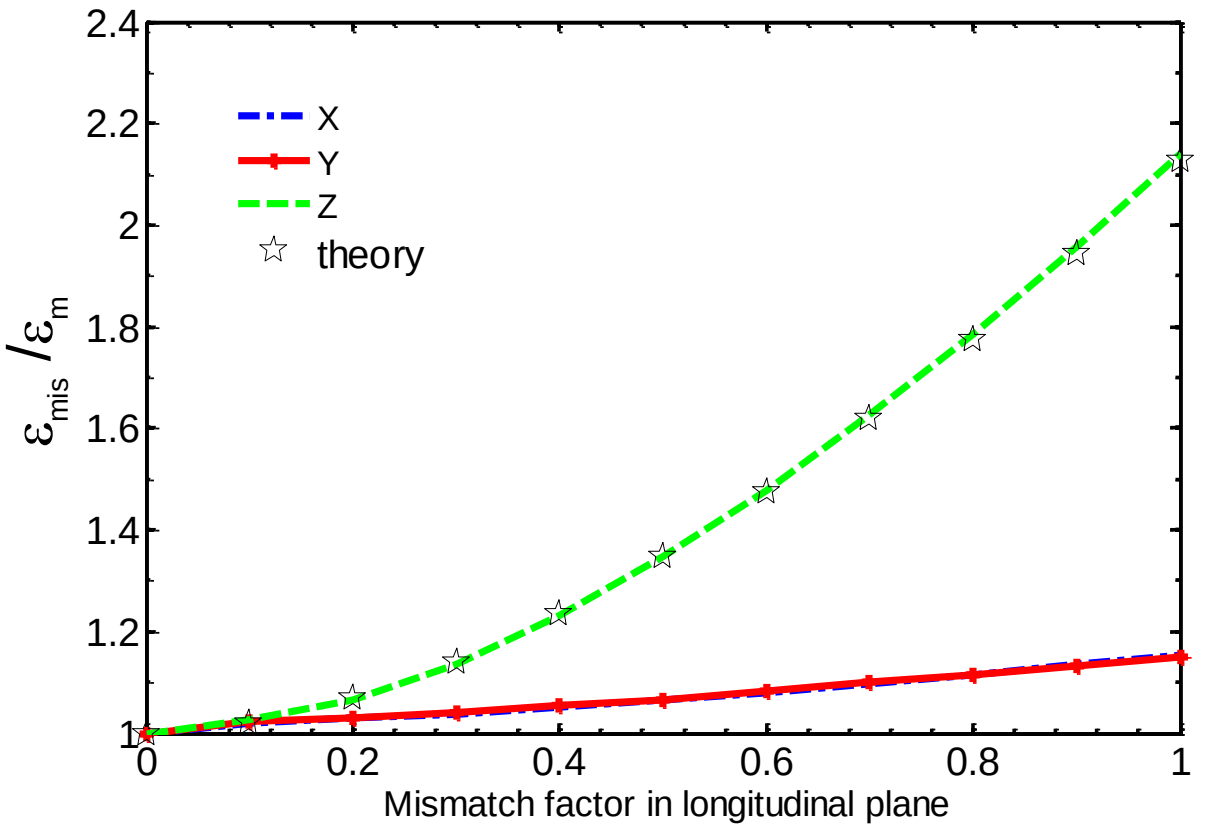

Fig. 2 RMS emittance growths at linac exit caused by mismatch in longitudinal plane.

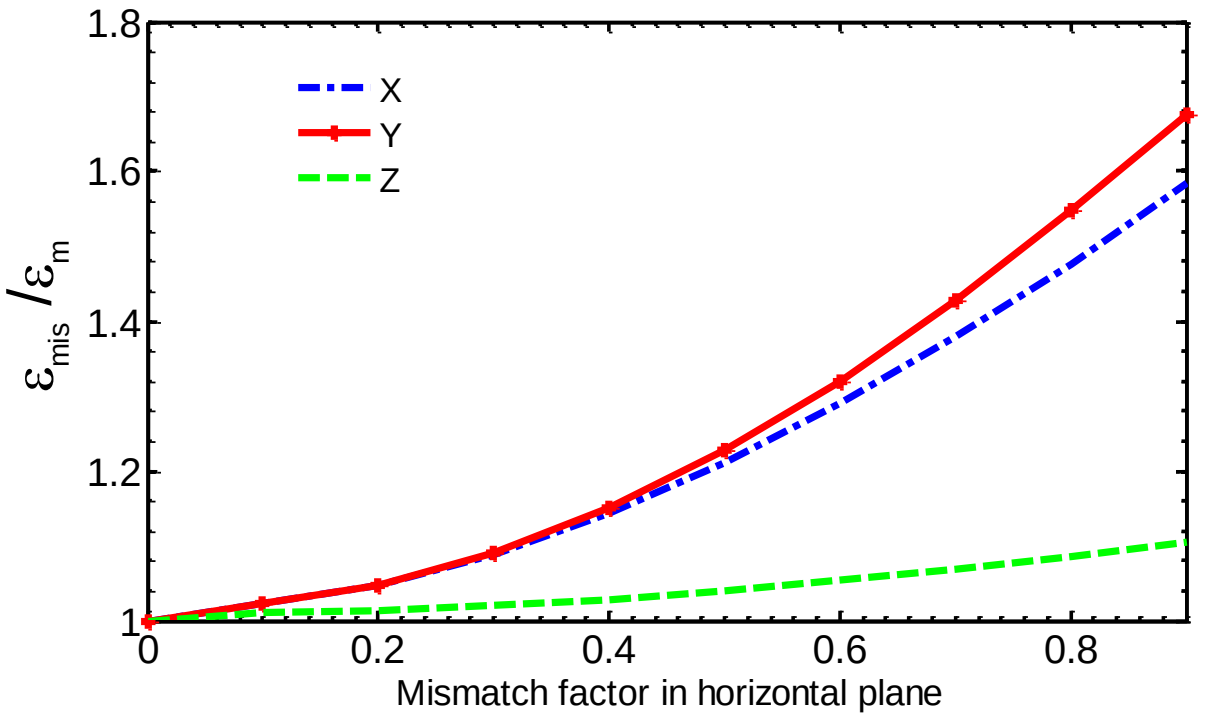

Fig. 3 RMS emittance growths at linac exit caused by mismatch in transverse plane.

Table 1: Emittance growth caused by mismatch in one plane

| Mismatch factor | | | Ex | Ey | Ez |
|---|---|---|---|---|---|
| x | y | z | % | % | % |
| 0 | 0 | 0 | 2.7 | 3.2 | 4.2 |
| +0.5 | 0 | 0 | 24.5 | 26.6 | 8.0 |
| 0 | 0 | +0.3 | 6.6 | 7.4 | 18.2 |

## 3.2 Envelope modes of mismatched bunched beams

The biggest frequency of the high eigenmodes for C-ADS accelerator is 155°, so there is no envelope instability. Figure 4 shows the parametric resonance between single particle and envelope oscillating. There are 1/2 parametric resonance for quadrupole mode and low mode and 1/3 parametric resonance for high mode in the C-ADS main linac. The low order resonances are the most dangerous ones and the high mode or low mode can excite a parametric resonance either in the transverse or longitudinal direction, so slow mode should be considered carefully.

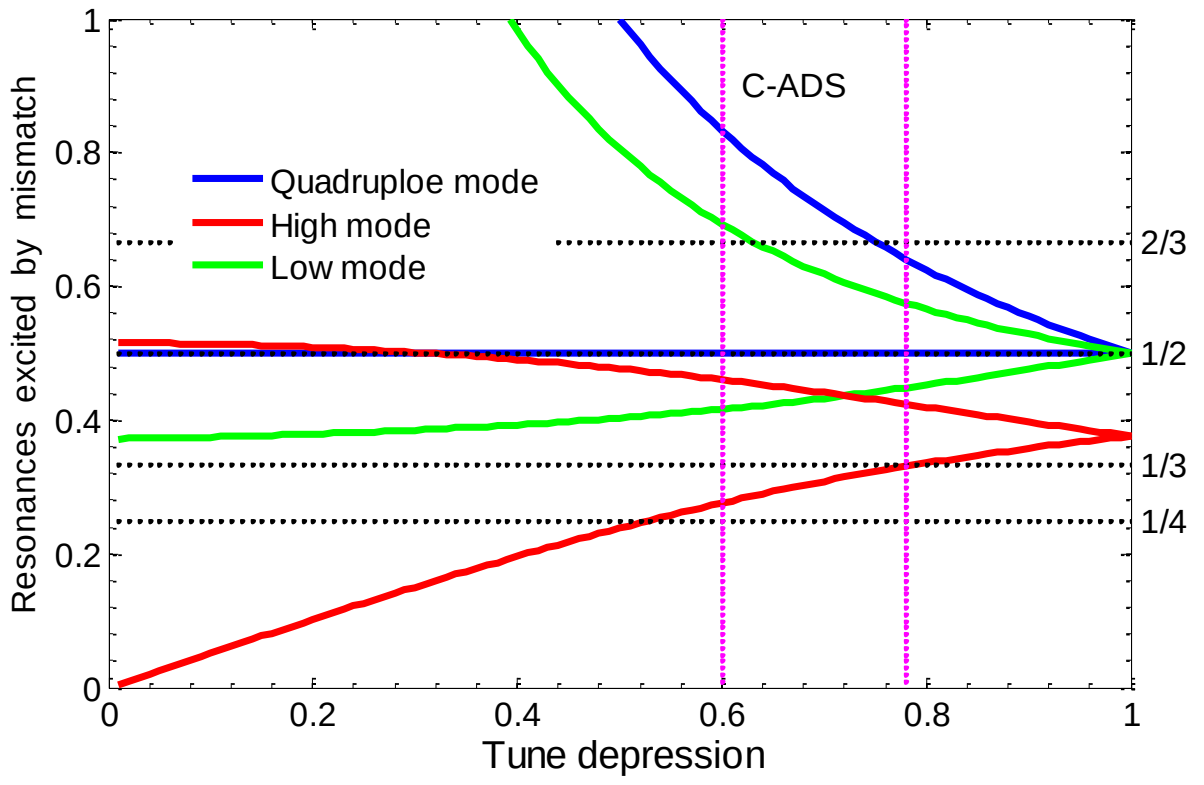

Fig. 4 Parametric resonance excited by mismatch.

For the quadrupole mode, one should keep quadrupole mode mismatch factor is smaller than 0.3, which means transverse mismatch factor is smaller than 0.4 to meet the requirement, and the rms emittance growth is shown in Fig 5. From the simulation results, one can see that emittance growth in horizontal direction is larger when quadrupole mode mismatch factor is larger than 0.5, which is because there are beam loss in vertical direction at the first few periods and because of the coupling effect of solenoid envelope oscillations shown in Fig 6 become stable fast. The $g_H$ of high mode is 0.28 for C-ADS main linac, which means that longitudinal mismatch factor is bigger than transverse mismatch factor. The emittance growths are shown in Fig 7, one should keep longitudinal mismatch factor within 0.3. However, the $g_L$ of low mode is -1.42, which means bigger transverse mismatch and the emittance growths are shown in Fig 8. Low mode mismatch factor should be smaller than 0.2 that means transverse mismatch factor smaller than 0.4. The difference of emittance growth in two transverse planes is also because of the difference of matched beam size at Spoke040 section. According to the simulation results shown in Table 3, transverse mismatch factor should be smaller than 0.4 and longitudinal mismatch smaller than 0.3 to meet the emittance growth requirement.

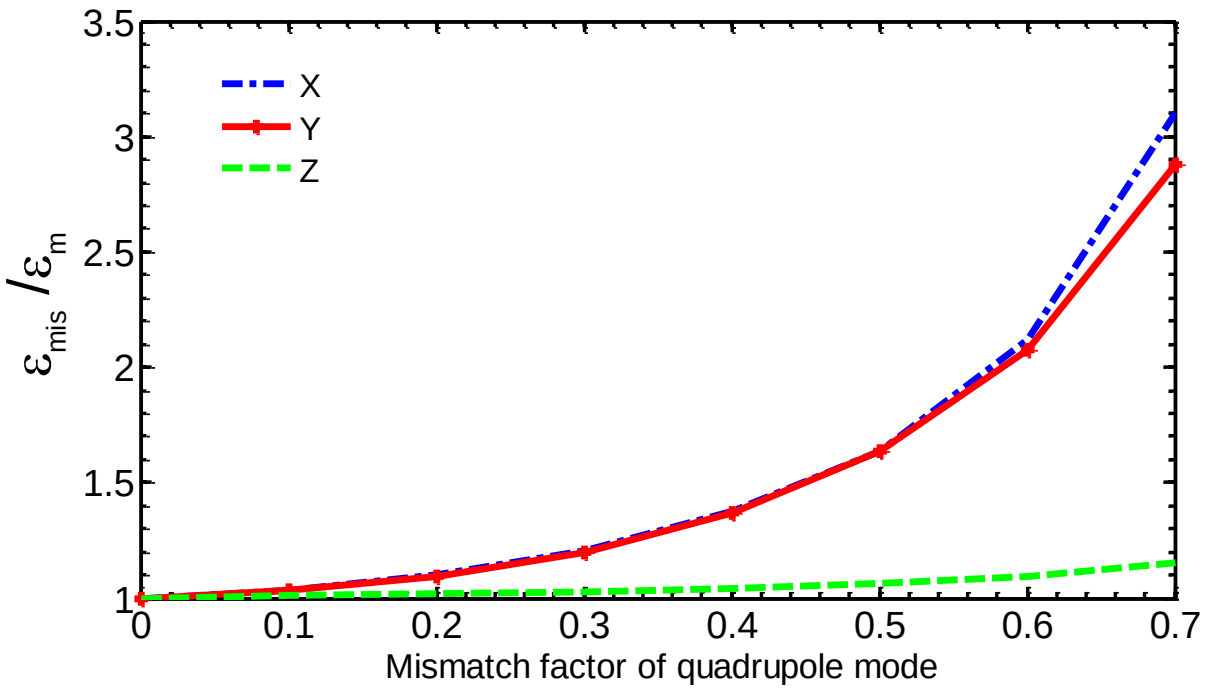

Fig. 5 RMS emittance growths at linac exit with quadrupole mode mismatch factor.

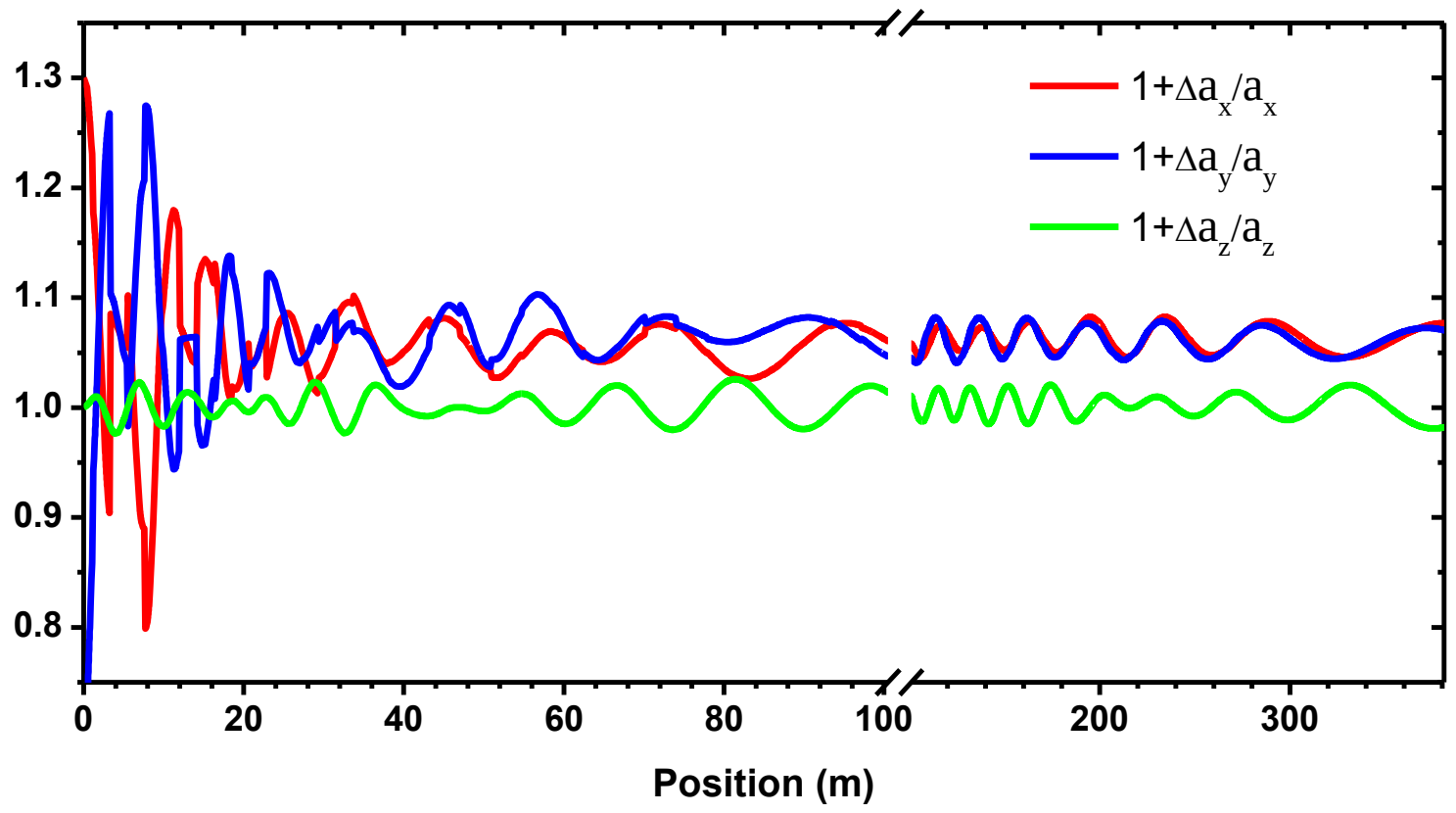

Fig. 6 Envelope oscillation with 0.3 mismatch quadrupole mode.

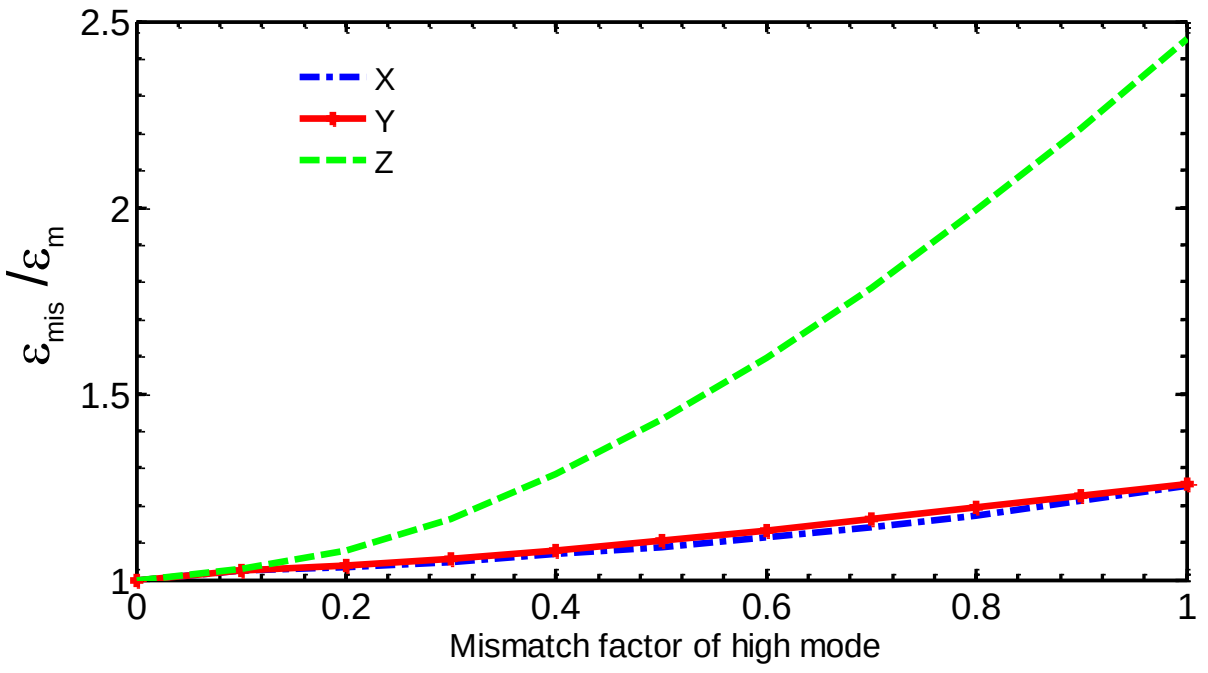

Fig. 7 RMS emittance growths at linac exit with high mode mismatch factor.

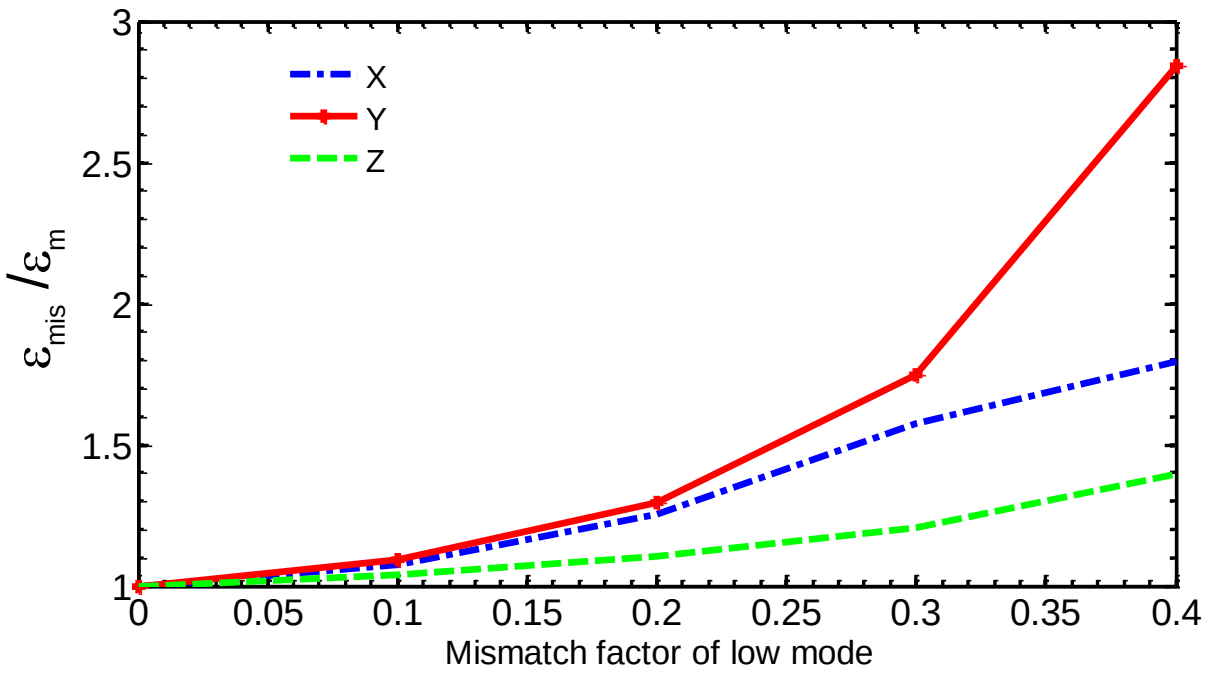

Fig. 8 RMS emittance growths at linac exit with low mode mismatch factor.

Table 2: Emittance growth caused by mismatch

| Mode | | Mismatch factor | | | Ex | Ey | Ez |
|---|---|---|---|---|---|---|---|
| | | x | y | z | % | % | % |
| | | 0 | 0 | 0 | 2.7 | 3.2 | 4.2 |
| Quad. | 0.3 | +0.3 | +0.43 | 0 | 23.9 | 23.3 | 6.9 |
| High | 0.3 | +0.08 | +0.08 | +0.3 | 7.7 | 9.0 | 21.2 |
| Low | 0.2 | -0.4 | -0.4 | +0.2 | 28.5 | 30.0 | 15.1 |

A mismatch with any amplitudes in transverse and longitudinal directions can lead to an excitation of the high and low mode simultaneously and the emittance growths are bigger. Figure 9 shows the envelope oscillation with 0.4 mismatch in transverse and 0.3 mismatch in longitudinal, the envelope oscillation correspond to the prediction.

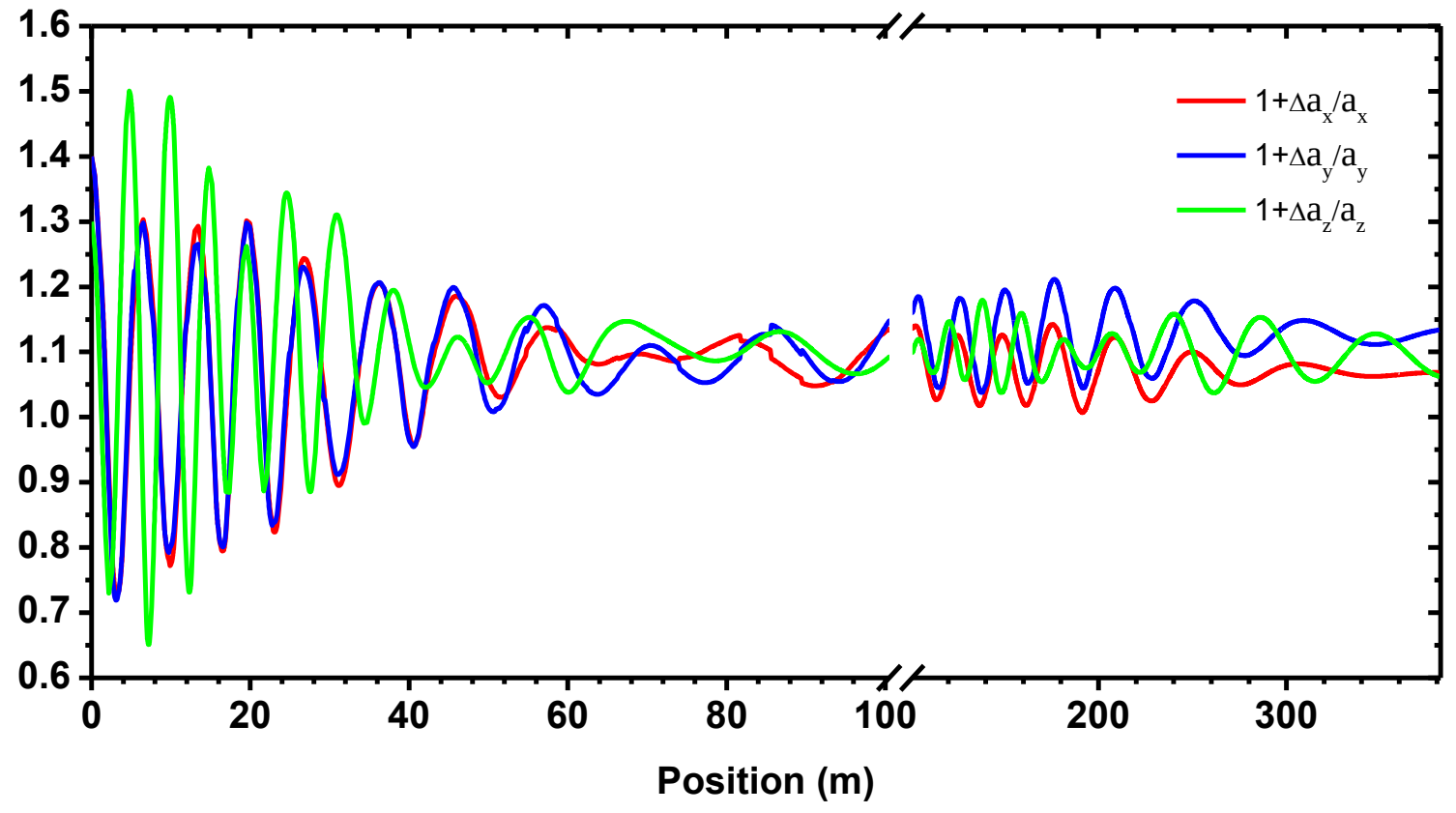

Fig. 9 Envelope oscillation with 0.4 mismatch in transverse and 0.3 mismatch in longitudinal.

## 4 Conclusions

Beam halo caused by mismatch is one major source of beam loss. The emittance growth and envelope oscillation caused by mismatch have been studied for the C-ADS accelerator. In this paper, the simulation results of the ADS main linac with beam mismatch are presented. The transverse mismatch factor should be smaller than 0.4 and longitudinal mismatch smaller than 0.3 to meet the emittance growth requirement.

*The authors would like to thank other colleagues in the ADS accelerator physics group for the discussions.*

# 中国 ADS 主加速器失匹配研究

孟才，唐靖宇，裴士伦，闫芳
粒子加速物理与技术重点实验室，中国科学院高能物理研究所，北京，100049

**摘要：** 中国 ADS 加速器是工作在连续波模式下的流强为 10mA、能量为 10MeV 的质子直线加速器，束流功率为 15 MW。由于束晕是束流丢失的一个主要原因，为了满足低束流损失和高稳定性的要求，研究束流失匹配导致的束晕是很重要的。为避免包络不稳定性，主加速器设计要求零流强下周期相移小于 90 度。本文介绍了主加速器段由束流失匹配导致的发射度增长和包络振荡。根据多粒子模拟的动力学结果，要保证发射度增长在一定范围内，横向、纵向失匹配因子分别小于 0.4 和 0.3。